\documentclass[floats,aps, amssymb,prd,onecolumn,secnumarabic,
tightenlines,nofootinbib]{revtex4}
\usepackage{amssymb}
\usepackage{graphics}
\usepackage{graphicx}
\usepackage{amsmath}
\usepackage{amsfonts}
\usepackage{bm}
\usepackage[colorlinks,linkcolor=red,anchorcolor=blue,citecolor=green]{hyperref}
\usepackage[export]{adjustbox}
\usepackage{ulem}   

\newcommand{\yq}[1]{#1}

\begin{document}

\title{
Quasinormal Modes in Noncommutative Schwarzschild Black Holes
}%

\author{Yaqi Zhao$^{1,2}$, Yifu Cai$^{1,2}$, S. Das$^3$, G. Lambiase$^{4,5}$, E.N. 
Saridakis$^{6,7,1,2}$, E.C. Vagenas$^{8}$}

\affiliation{$^1$Deep Space Exploration Laboratory/School of Physical Sciences, University of Science and Technology of China, Hefei, Anhui 230026, China}

\affiliation{$^2$CAS Key Laboratory for Researches in Galaxies and Cosmology/Department of Astronomy, School of Astronomy and Space Science, University of Science and Technology of China, Hefei, Anhui 230026, China}

\affiliation{
$^3$Theoretical Physics Group and Quantum Alberta, Department of Physics and Astronomy,
University of Lethbridge, 4401 University Drive, Lethbridge, Alberta, T1K 3M4, Canada. }

\affiliation{$^4$Dipartimento di Fisica ''E.R. Caianiello'' Universit\`a di Salerno, I-84084 Fisciano (Sa), Italy,}

\affiliation{$^5$INFN - Gruppo Collegato di Salerno, Italy.}

\affiliation{$^6$ National Observatory of Athens, Lofos Nymfon, 11852 Athens, 
Greece}

\affiliation{$^7$Departamento de Matem\'{a}ticas, Universidad Cat\'{o}lica del 
Norte, 
Avda. Angamos 0610, Casilla 1280 Antofagasta, Chile}

\affiliation{$^8$Department of Physics, College of Science, Kuwait University, Sabah Al Salem University City, P.O. Box 2544, Safat 1320, Kuwait}

\def\be{\begin{equation}}
\def\ee{\end{equation}}
\def\al{\alpha}
\def\bea{\begin{eqnarray}}
\def\eea{\end{eqnarray}}

\begin{abstract}
\par\noindent
We investigate the quasinormal modes of a massless scalar field in  a Schwarzschild black hole, which is deformed due to noncommutative corrections. 
We introduce the  deformed Schwarzschild black hole solution, which depends on the noncommutative parameter $\Theta$.  We then extract the master equation as a  Schr\"odinger-like equation, giving the explicit expression of the effective potential which is modified due to the noncommutative corrections. After that,  we solve the master equation numerically. The significance of these results is twofold. Firstly, our results can be related to the detection of gravitational waves by the near future gravitational wave detectors, such as LISA, which will have a significantly increased accuracy. In particular, these observed gravitational waves produced by binary strong gravitational systems have oscillating modes which can provide valuable information.  Secondly, our results can serve as an additional tool to test the predictions of GR, as well as to examine the possible detection of this kind of gravitational corrections.

\end{abstract}


\maketitle
%
%
%
%
%
\section{Introduction}
\label{sec: introduction}
\setcounter{equation}{0}
%
%
%
\par\noindent
In a seminal paper by Snyder \cite{Snyder:1946qz}, there appeared for the first time the idea that spacetime can be described in noncommutative (NC) frameworks. Such a consideration, and more generally the interest for the NC physics, received a renewed interest later on due to  the discovery of the Seiberg-Witten map, which essentially relates the NC theories to the commutative gauge theories \cite{Seiberg:1999vs}. The latter has played a relevant role in the understanding of NC physics on a fundamental level, that is the spacetime symmetries and
%
%
the unitary properties of these theories \cite{Alvarez-Gaume:2006qrw,Carroll:2001ws,Chaichian:2002vw,Doplicher:2001qt,Doplicher:1994tu,Bahns:2002vm,Iorio:2001qy,Jackiw:2001jb, Banerjee:2003vc, Banerjee:2004rs}, and the possible experimental signatures \cite{Amelino-Camelia:2002onw,Guralnik:2001ax,Cai:2001az,Douglas:2001ba,Szabo:2001kg}  (see also \cite{Alvarez-Gaume:2006qrw,Carroll:2001ws,Chaichian:2002vw,Doplicher:2001qt,Doplicher:1994tu,Bahns:2002vm,Iorio:2001qy,Jackiw:1978ar,Jackiw:2001jb,Amelino-Camelia:2002onw,Guralnik:2001ax,Cai:2001az,Bichl:2001yf,Grimstrup:2002xs} for the violation of spacetime symmetries and \cite{Doplicher:2001qt,Doplicher:1994tu,Douglas:2001ba,[OMMY]} for the full Lorentz invariance). Furthermore, in the framework of string theories, the low-energy limit yields in a natural way a quantized structure of spacetime \cite{Ardalan:1998ce,Seiberg:1999vs,Douglas:2001ba,Szabo:2001kg}, while at high-energy scales the noncommutativity of spacetime may lead to a deep insight on its quantum nature \cite{Seiberg:1999vs,Doplicher:2001qt,Doplicher:1994tu, Banerjee:2009gr}.

In the NC frameworks, one defines the fields over phase space in which the ordinary product of fields is replaced by the Moyal product. Due to the Seiberg-Witten map, this theory turns out to be equivalent to commutative gauge theories in which the fields are expanded in terms of a NC parameter \cite{Dayi:2002za,Ghosh:2003ka,Chakraborty:2003cp,Ghosh:2004ee,Mukherjee:2004cb,Saha:2006qt,Calmet:2005qm,Chaichian:2001py,Chaichian:2004yw,Calmet:2001na,Aschieri:2002mc, Mukherjee:2007fa}. In this respect, the formulation of gravity as a commutative equivalent gauge theory turns out to be very promising (for details, see \cite{Chamseddine:2000si,Bonora:2000td,Jurco:2000ja,Chaichian:2004za,Chaichian:2004yh,Aschieri:2005yw,Alvarez-Gaume:2006qrw, Calmet:2005qm,Calmet:2006iz,Kobakhidze:2006kb,Chaichian:2006we,Chaichian:2007we,Chaichian:2006ht,Manolakos:2019fle, Manolakos:2022ihc}).

On the other hand, recently the study of the quasinormal modes (QNMs) \cite{Nollert:1999ji, Konoplya:2011qq, Isi:2021iql} has generated a lot of interest in the scientific community \cite{Cai:2015fia, Cardoso:2019mqo, McManus:2019ulj, Guo:2021enm, Ejlli:2019bqj, Pantig:2022gih}, since they allow for the investigation of gravity in the strong-field regime. This is certainly possible due to experiments with increasing accuracy, such as the Event Horizon Telescope (EHT), which has captured the first image of a black hole \cite{EventHorizonTelescope:2019dse, EventHorizonTelescope:2019ggy, EventHorizonTelescope:2020qrl}, and the LIGO-Virgo collaboration which has detected the first gravitational wave signal \cite{LIGOScientific:2016aoc,  LIGOScientific:2016lio}. These observations opened the possibility of studying black hole features  near the event horizon, as well as of probing modified gravity theories \cite{Cai:2018rzd, Yan:2019hxx, Li:2021mzq,Jusufi:2021fek}, or quantum gravity corrections \cite{ Dreyer:2002vy, Lewandowski:2022zce}, which is the subject of this paper.
%
%
More specifically, since QNMs represent characteristic modes of the perturbation equations in a given gravitational background \cite{Zerilli:1970se, Chandrasekhar:1975zza, Kokkotas:1999bd, Berti:2009kk, Hatsuda:2021gtn}, they provide information on the geometry of spacetime. Such a feature highlights the important role of QNMs in connection with the physics of gravitational waves (GWs). In fact, on one hand GWs allow us to make observations in order to test general relativity (GR) \cite{Dreyer:2003bv, Berti:2005ys, Isi:2019aib}, and on the other hand the predictions of QNMs properties allow us to constrain the gravitational theories beyond GR \cite{Cano:2021myl, Wang:2004bv, Blazquez-Salcedo:2016enn, Franciolini:2018uyq,Becar:2019hwk,Aragon:2020xtm, Liu:2020qia, Karakasis:2021tqx, Gonzalez:2022upu,Ishibashi:2003ap, Zhao:2022gxl, Chowdhury:2022zqg}. 

The aim of this paper is to investigate the  QNMs in the case of black hole solutions in NC gravity. In particular, we refer to the {\it deformed} Schwarzschild solution, where the corrections are induced by the NC parameter $\Theta$. For this metric, we compute the  corresponding QNM frequency of a massless scalar field. 
The noncommutativity of gravity is induced by a NC coordinate \yq{algebra} given by
\begin{equation}\label{nccoord}
    [x^\mu, x^\nu] = i\Theta^{\mu\nu},
\end{equation}
where the (antisymmetric) tensor $\Theta^{\mu\nu}$  is a $c$-number (here the Greek indices are used for the spacetime coordinates $\mu,\nu= 0,\ldots, 3$) and accounts for the {\it degree of quantum fuzziness} of spacetime. At this point it should be noted that  different approaches have been proposed in which the NC coordinates occur, for instance in the $q$-deformed theories \cite{Madore:2000en}. The NC parameter $\Theta^{\mu\nu}$ has been constrained in several frameworks: in low-energy measurements \cite{Mocioiu:2000ip,Chaichian:2000si,Chaichian:2002ew,Joby:2014oee}, in Lorentz symmetry breaking \cite{Carroll:2001ws,Calmet:2004dn}, in cosmology and physics of the primordial Universe \cite{Joby:2014oee,Calmet:2015fma,Lambiase:2016dpc,Addazi:2021xuf}, and in gravitational physics 
\cite{Chamseddine:2000si,Aschieri:2005yw,Calmet:2005qm,Aschieri:2005zs,Kobakhidze:2006kb,Szabo:2006wx,Calmet:2006iz,Mukherjee:2006nd,Kobakhidze:2016cqh,Kobakhidze:2007jn,Kanazawa:2019llj, CANTATA:2021ktz}. 
It is noteworthy that in the aforesaid models the NC corrections appear to second order in $\Theta$. In TABLE~\ref{table1}, we report some bounds on $\Theta^{\mu\nu}$.
\begin{table}[!ht]
    \label{table1}
	\begin{center}
	\begin{tabular}{|c|c|c|}
	\hline
    \text{Bounds on} $\Theta$ & \text{Physical framework} & \text{Refs.} \\ \hline 
    \hline
    $\Theta< (1 \text{TeV})^{-2}$ & Nucleus wave function & \cite{Carroll:2001ws,Calmet:2004dn} \\
    $\Theta<10^{-8}$GeV$^{-2}$ & Lamb shift corrections & \cite{Chaichian:2000si,Chaichian:2002ew} \\
    $\Theta<  10^{-7}$GeV$^{-2}$ & CMB physics  & \cite{Joby:2014oee} \\
    $|\Theta|<10^{-11}$GeV$^{-1}$ & Generalized uncertainty principle (GUP) & \cite{Kanazawa:2019llj} \\
    $|\Theta|< 8.4\times 10^{-38}$GeV$^{-2}$ & GW signal detected by LIGO/Virgo collaboration & \cite{Kobakhidze:2016cqh} \\
 	\hline 
	\end{tabular}
	\caption{Bounds on the NC parameter  $\Theta$ (or $|\Theta|$) inferred in different experiments, where $\Theta$ refers to different components of $\Theta^{\mu\nu}$ (see the corresponding references). The different units of $\Theta$ in the GUP case arises from the fact that  there the bounds have been inferred in spherical coordinates and therefore $[\Theta]=[\Theta^{r\theta}]=\text{GeV}^{-1}$, while in Cartesian coordinates $[\Theta]= \text{GeV}^{-2}$.}
	\end{center}
\end{table}

\par\noindent
The rest of the paper is organized as follows. In Section \ref{BHTheta}, we review the NC gravity and present the deformed Schwarzschild black hole solutions which are $\Theta$-depending. In
Section \ref{NCgeometryQNMs}, we derive the form of the Schr\"odinger-like equation. In Section \ref{sec: QNM calculation}, we numerically solve the Schr\"odinger-like equation, and get the time evolution of the dominant mode. In Section \ref{Conclusions}, we present and discuss our conclusions.
%
%
%
\section{$\Theta^{\mu\nu}$- Schwarzschild black holes}
\label{BHTheta}
%
%
%
\par\noindent
In this Section we briefly review the NC black hole solutions  following the analysis of Ref. \cite{Chaichian:2007we}. The NC corrections to the Schwarzschild black hole geometry, which are a $\Theta$-expansion, are investigated (for other solutions of the deformed Einstein field equations the reader could see \cite{Chaichian:2006we,Chaichian:2007we,Chaichian:2001py,Chaichian:2004yw,Kobakhidze:2016cqh,Kobakhidze:2007jn} and references therein). The important point is that the Schwarzschild black hole solution is {\it exact}~
\footnote{\yq{Note that having an exact solution gives one the advantage of having full control over the correction terms. This is especially useful when one wants to do an analysis in which it is important to know the order of each term, so that at a suitable point in the analysis, lower order terms can be discarded, e.g., in the analysis of QNMs which we present in Section \ref{sec: QNM calculation}. On the other hand, having an approximate solution and introducing correction terms, e.g., due to NC gravity, does not give one full control of the terms, to clearly identify terms of higher and lower orders and the sources of those term. This potentially reduces the value of the results obtained in terms of their correctness and accuracy. }}. 
One starts by writing the deformed metric in terms of the tetrad fields, namely
\begin{equation}\label{deformedmetric}
   {\hat g}_{\mu\nu}(x, \Theta)=\frac{1}{2}\left({\hat e}_\mu^a \ast  {\hat e}_\nu^{b \, \dagger}+{\hat e}_\mu^b \ast {\hat e}_\nu^{a \, \dagger}\right)\eta_{ab},
\end{equation}
where the ``${}^\dagger$" denotes the complex conjugation, the Latin indices are used for the tangent space basis $a, b = 0, \dots , 3$, and the ``$\ast$" stands for the Moyal product which is defined as $\phi(x) \ast \chi(x) \equiv e^{\frac{i}{2} \Theta^{\mu\nu}\partial_{x^\mu}\partial_{y^\nu}}\phi(x) \chi(y)\Big|_{y\to x}$. The tetrads, as gauge fields, are expanded in terms of the $\Theta^{\nu\rho}$-parameter as
\begin{equation}\label{tetradNC}
 {\hat e}^a_\mu (x, \Theta)=e^a_\mu(x)-i\Theta^{\nu\rho}e^a_{\mu\nu\rho}(x)+\Theta^{\nu\rho}\Theta^{\alpha\sigma}e^{a}_{\mu\nu\rho\alpha\sigma}(x)+{\cal O}(\Theta^3),
\end{equation}
with 
\[
e^a_{\mu\nu\rho}(x)=\frac{1}{4}[\omega^{ac}_\nu \partial_\rho e^d_\mu+(\partial_\rho \omega_\mu^{ac}+R_{\rho\mu}^{ac})e^d_\nu]\eta_{cd},
\]
and similar expressions  exist for the other terms of the expansion \cite{Chaichian:2007we}. In the above equations, $R_{\rho\mu}^{ac}$ is the curvature tensor and $\omega_\mu^{ac}$ is the connection.

\begin{figure}[h]
    \centering
    \includegraphics[width=0.2\textwidth]{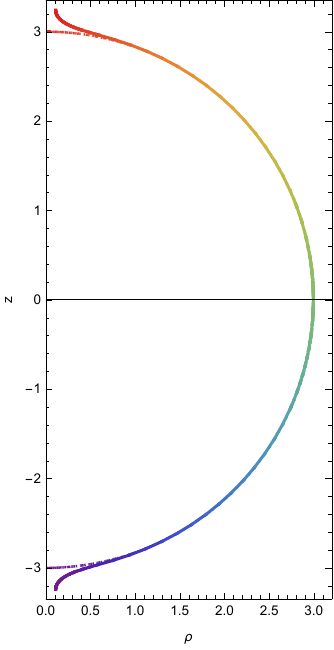}
    \includegraphics[width=0.36\textwidth,frame]{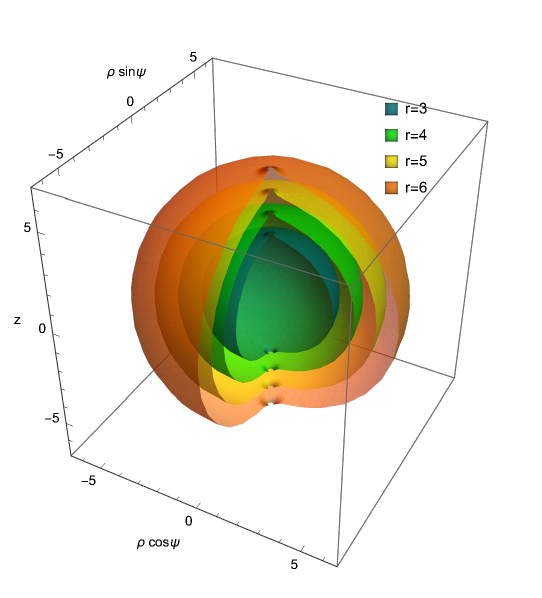}
    \includegraphics[width=0.36\textwidth]{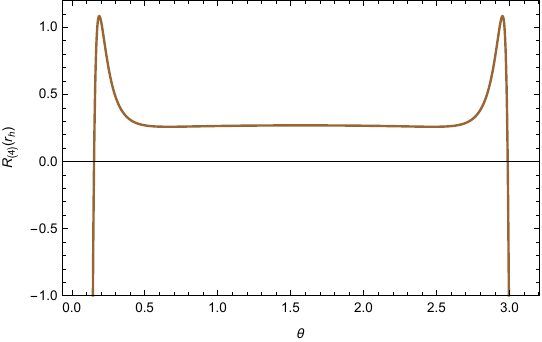}
    \caption{\it{The left panel shows the embedded shape $(\rho(\theta),z(\theta))$ for a constant $\varphi$ and $r=3$. The thick curve depicts the case of the NC metric, while, given for comparison,  the dashed curve corresponds to the sphere geometry. The color of the curve indicates the    $\theta$ value. The middle panel shows the $3$-dimensional shape of the constant $r$ slides. The color of the surface indicates the value of $r$. The parameter choice is $\Theta^2=0.2$ for both plots, and we have adopted $M=1$. Besides, the right panel shows the first order curvature scalar at horizon radius $R_{(4)}(r_h)$.}}
    \label{fig: shell}
\end{figure}

Following the quantization procedure of Refs. \cite{Chaichian:2006ht,Wang:2008ut,Wang:2009ju,Sun:2010nas} for the Schwarzschild black hole metric, one needs to fix the Moyal algebra. In terms of the spherical coordinates $x^\mu = (t, r, \theta, \phi)$, the algebra is deformed as
\begin{equation}
    \label{ThetaCh}
    \Theta^{\mu\nu}= \Theta 
    \begin{pmatrix}
    0 & 0 & 1 & 0 \\
    0 & 0 & 0 & 0 \\ -1 & 0 & 0 & 0 \\ 0 & 0 & 0 & 0 
    \end{pmatrix},
\end{equation}
where $\Theta$ is the deformation parameter, which as discussed in the Introduction, gives rise to the simplest model of a NC spacetime.
%
The non-deformed Schwarzschild black hole geometry is described by the line element $ds^2=g_{\mu\nu}^{(S)}dx^\mu dx^\nu$, with 
\begin{equation}
    \label{metricdef}
    g_{\mu\nu}^{(S)}=\text{diag}\left(-(1-\frac{\alpha}{r}), (1-\frac{\alpha}{r})^{-1}, r^2, r^2\sin^2\theta\right)\,, \quad \alpha \equiv 2GM,
\end{equation}
where $M$ is the mass of the gravitational source. Hence, the corresponding vierbein fields are 
\begin{equation}\label{tetraddefs}
    e_\mu^0=\left(1-\frac{\alpha}{r}, 0, 0, 0\right), \quad  e_\mu^1=\left(0, (1-\frac{\alpha}{r})^{-1}, 0, 0\right)\,, 
    \quad e_\mu^2=(0, 0, r, 0)\,, \quad
    e_\mu^3=(0, 0, 0, r\sin\theta)~. 
\end{equation}
Furthermore, the components of the $\Theta$-Schwarzschild black hole metric, according to (\ref{deformedmetric}), are   given by
\begin{equation}\label{deformedS}
  {\hat g}_{\mu\nu}=g_{\mu\nu}^{(S)}+h_{\mu\nu}^{(NC)}
\end{equation}
with
\begin{eqnarray}
  h_{00}^{(NC)} &=& -\frac{\alpha(8r-11\alpha)}{16r^4}\Theta^2+{\cal O}(\Theta^4)\,, \label{h00NC} \\
  h_{rr}^{(NC)} &=& -\frac{\alpha(4r-3\alpha)}{16r^2(r-\alpha)^2}\Theta^2 + {\cal O}(\Theta^4)\,, \label{hrrNC} \\
  h_{\theta\theta}^{(NC)} &=& \frac{2r^2-17\alpha (r-\alpha)}{32r(r-\alpha)}\Theta^2+{\cal O}(\Theta^4)\,, \label{thetathetaNC} \\
  h_{\phi\phi}^{(NC)} &=& \frac{(r^2+\alpha r-\alpha^2)\cos^2\theta-\alpha 
  (2r-\alpha)}{16r(r-\alpha)}\Theta^2+{\cal O}(\Theta^4),  \label{hphiphiNC}
\end{eqnarray}
where $h_{\mu\nu}^{(NC)}$  quantifies the NC corrections to the Schwarzschild black hole metric. \yq{To display the geometry of this metric, we have embedded its 2-dim slice $(\theta, \varphi)$ with fixed $t = 0$ and $r = \text{const}$ into a 3-dim flat space with cylindrical coordinate system $(\rho,\psi,z)$, as shown in FIG.~\ref{fig: shell}.} 
%
%
%
\par\noindent
At this point, a number of comments are in order. First,  the standard Schwarzschild black hole solution is recovered in the limit $\Theta\to 0$, as anticipated. Second, as mentioned in the Introduction, the corrections enter in the metric as $\Theta^2$, that is they are of second order in the deformation parameter $\Theta$ (this is a general aspect of NC gravity \cite{Chamseddine:2000si,Aschieri:2005yw,Calmet:2005qm,Aschieri:2005zs,Kobakhidze:2006kb,Szabo:2006wx,Calmet:2006iz,Mukherjee:2006nd,Kobakhidze:2016cqh,Kobakhidze:2007jn}). Third, it is worth of noting that since we are using spherical coordinates, the dimension of $\Theta$ are $[\Theta]^2=[L]^2=[M]^{-2}$ \cite{Chaichian:2007we,Sun:2010nas,Wang:2008ut,Chaichian:2007dr,Singh:2017bwj,Alavi:2013lum}, contrary to the standard canonical quantization, which is in Cartesian coordinates, where the NC parameter $\Theta^2$ has dimensions $[L]^4$.

%
%
%
\section{Quasinormal Modes in Noncommutative black hole spacetime}
\label{NCgeometryQNMs}
%
%
%
\par\noindent
Now we proceed to the investigation of the QNMs in the deformed, due to noncommutativity, Schwarzschild black holes. We start with the derivation of  the effective potential by writing the Klein-Gordon field equation in a Schr\"odinger-like form. For this reason, we adopt the analysis of Ref. \cite{Chen:2022ynz} where a general formalism has been developed through ``projection method".

\par\noindent
The deformed Schwarzschild black hole metric is parameterized as 
%
%
%
\begin{eqnarray}
    g_{tt}&=& -\left(1-\frac{r_h}{r}\right)\left(1+\epsilon A_j(r)\cos^j{\theta}+\mathcal{O}(\epsilon^2)\right)\,,\label{gtt}\\
    g_{rr}&=& \left(1-\frac{r_h}{r}\right)^{-1}\left(1+\epsilon B_j(r)\cos^j{\theta}+\mathcal{O}(\epsilon^2)\right)\,,\\
    g_{\theta\theta}&=& r^2\left(1+\epsilon C_j(r)\cos^j{\theta}\right)\,,\label{gthetatheta}\\
    g_{\varphi\varphi}&=& r^2\sin^2\theta\left(1+\epsilon D_j(r)\cos^j{\theta}\right)\,,\label{gvarphivarphi}\\
    g_{tr}&=& \epsilon a_j(r)\cos^j{\theta}\,,\qquad g_{t\theta}=\epsilon b_j(r)\cos^j{\theta}\,,\label{gtr}\\
    g_{r\theta}&=& \epsilon c_j(r)\cos^j{\theta}\,,\qquad g_{r\varphi}=\epsilon d_j(r)\cos^j{\theta}\,,\label{rtheta}\\
    g_{\theta\varphi}&=& \epsilon e_j(r)\cos^j{\theta}, \label{gthetavarphi}
\end{eqnarray}
where the index $j$ stands for summations running upward from $j=0$\yq{, and $r_h$ is the position of the event horizon of the black hole defined as the largest positive zero point of function $1/g_{rr}$}. 
By comparing (\ref{gtt})-(\ref{gthetavarphi}) with (\ref{h00NC})-(\ref{hphiphiNC}), one obtains 
\begin{eqnarray}
    \epsilon &=& \Theta^2\,, \label{epsilonTheta} \\
    A_0(r) &=& \frac{3r^2+3r\alpha+11\alpha^2}{16r^3\alpha}\,, \label{A0}\\
    B_0(r) &=& -\frac{3r^3+4r\alpha^2-3\alpha^3}{16r^3(r-\alpha)\alpha}\,, \label{B0} \\
    C_0(r) &=& \frac{2r^2-17\alpha (r-\alpha)}{32r^3(r-\alpha)}\,, \label{C0} \\
    D_0(r) &=& -\frac{\alpha(2r-\alpha)}{16r^3(r-\alpha)}\,, \label{D0} \\
    D_j(r) &=& \frac{1+(-1)^j}{32r^2} \quad\quad (\text{for}\; j>0) \;, \label{Dj} \\
    a_j(r)&=& b_j(r)=c_j(r)=d_j(r)=e_j(r)=0~, \label{ajdj=0}
\end{eqnarray}
and \yq{by solving the equation $1/g_{rr}=0$, one acquires the exact solution for $r_h$}
\begin{equation}
\tiny
\begin{aligned}
    r_h=
&
    \frac{1}{4} \left(\left(\sqrt[3]{-22 \alpha ^4 \epsilon +8 \alpha ^2 \epsilon ^2-2 \sqrt{\alpha ^4 \epsilon ^2 \left(121 \alpha ^4+344 \alpha ^2 \epsilon +16 \epsilon ^2\right)}}+\alpha ^2 \left(1-\frac{6\ 2^{2/3} \epsilon }{\sqrt[3]{-11 \alpha ^4 \epsilon +4 \alpha ^2 \epsilon ^2-\sqrt{\alpha ^4 \epsilon ^2 \left(121 \alpha ^4+344 \alpha ^2 \epsilon +16 \epsilon ^2\right)}}}\right)\right)^{1/2}\right.
    \\
&
    \left.+2 \left(\frac{\alpha ^2}{2}-\frac{\sqrt[3]{-11 \alpha ^4 \epsilon +4 \alpha ^2 \epsilon ^2-\sqrt{\alpha ^4 \epsilon ^2 \left(121 \alpha ^4+344 \alpha ^2 \epsilon +16 \epsilon ^2\right)}}}{2\ 2^{2/3}}+\frac{3 \alpha ^2 \epsilon }{\sqrt[3]{-22 \alpha ^4 \epsilon +8 \alpha ^2 \epsilon ^2-2 \sqrt{\alpha ^4 \epsilon ^2 \left(121 \alpha ^4+344 \alpha ^2 \epsilon +16 \epsilon ^2\right)}}}\right.\right.
    \\
&
    \left.\left.+\frac{\alpha ^3-4 \alpha  \epsilon }{2 \sqrt{\sqrt[3]{-22 \alpha ^4 \epsilon +8 \alpha ^2 \epsilon ^2-2 \sqrt{\alpha ^4 \epsilon ^2 \left(121 \alpha ^4+344 \alpha ^2 \epsilon +16 \epsilon ^2\right)}}+\alpha ^2 \left(1-\frac{6\ 2^{2/3} \epsilon }{\sqrt[3]{-11 \alpha ^4 \epsilon +4 \alpha ^2 \epsilon ^2-\sqrt{\alpha ^4 \epsilon ^2 \left(121 \alpha ^4+344 \alpha ^2 \epsilon +16 \epsilon ^2\right)}}}\right)}}\right)^{1/2}+\alpha \right)
\end{aligned}.
\normalsize
\end{equation}
In addition, \yq{expanding the above expression} to the $\Theta^2$ order, we have
\begin{align}
    \label{eq: rh}
    r_h=\alpha+\frac{3\Theta^2}{16\alpha}+\mathcal{O}(\Theta^4)~.
\end{align}
We have calculated the curvature scalar at the horizon radius \eqref{eq: rh}, showing
\begin{equation}
    R_{(4)}(r_h)=-\frac{9 \left(78608 \csc ^2(\theta ) \left(6936 \csc ^4(\theta )-322116 \csc ^2(\theta )+1543913\right)-1604119805285\right)}{3163759443968 \alpha ^4} \epsilon +\mathcal{O}(\epsilon^2).
\end{equation}
The graph for the first order coefficient of this expressioin is shown in the right panel in FIG. \ref{fig: shell}. We have set $M=1$ in this plot. From the plot, we can see that this black hole has a horizon and not a singular point at $r=r_h$, different from the Chern-Simons scenario~\cite{Nakashi:2020phm}.

%
%
%
\par\noindent
Next, we consider massless scalar waves propagating in the deformed Schwarzschild black hole spacetime, with equation of motion of the form
\begin{equation}\label{KGequationmaster}
    \Box\psi=0~.
\end{equation}
Exploiting the fact that there are two Killing vectors, i.e., $\partial_t$ and $\partial_\varphi$, this equation can be decomposed through
\begin{equation}\label{BoxpsiDecomp}
    \Box\psi=\int_{-\infty}^\infty d\omega \sum_{m=-\infty}^\infty 
    e^{i(m\varphi-\omega t)}\mathcal{D}_{m,\omega}^2\psi_{m,\omega}(r,\theta),
\end{equation}
where the Fourier modes of the wave function, i.e.,  $\psi_{m,\omega}$, satisfy the equation
\begin{equation}
    \mathcal{D}_{m,\omega}^2\psi_{m,\omega}=0,
\end{equation}
with $m$ the azimuthal number and $\omega$ the mode frequency. For the master equation we consider, the variable separation for ``quantum number" $m$ is always precisely achievable, while in general we can't do this precisely for the ``quantum number" $l$. In order to achieve a separation for the ``quantum number" $l$ at the dominant order of $\epsilon$ (or $\Theta^2$ here), we need to follow the ``projection method"~\cite{Chen:2022ynz, Cano:2020cao} as described below. The operator $\mathcal{D}_{m,\omega}^2$ can be written (up to first order in $\epsilon$) as
\begin{equation}\label{KGop}
    \mathcal{D}_{m,\omega}^2=\mathcal{D}_{(0)m,\omega}^2+\epsilon\mathcal{D}_{(1)m,\omega}^2 ~.
\end{equation}
The zeroth order operator, i.e., $\mathcal{D}_{(0)m,\omega}^2$, is given by
\begin{equation}\label{KGop0}
    \mathcal{D}^2_{(0)m,\omega}=-\left[\omega^2-\frac{m^2 {\cal F}(r)}{r^2\sin^2{\theta}}
    \right]-\frac{{\cal F}(r)}{r^2}\partial_{r}\left[r^2 {\cal F}(r)\partial_{r}\right]-\frac{{\cal F}(r)}
    {r^2\sin{\theta}}\partial_{\theta}\left(\sin{\theta}\partial_{\theta}\right),
\end{equation}
with ${\cal F}(r)\equiv 1-\frac{r_h}{r}$, while the first order operator, i.e., $\mathcal{D}_{(1)m,\omega}^2$, is presented in (\ref{KGop1}) of Appendix \ref{AppendixA} \cite{Chen:2022ynz}. In addition,  the Fourier modes of the wavefunction, i.e., $\psi_{m,\omega}$,  can be expanded as
\begin{equation}\label{exppsiRlm}
    \psi_{m,\omega}=\sum_{l'=|m|}^\infty P_{l'}^m(x)R_{l',m}(r),
\end{equation}
where $x=\cos\theta$ and the Legendre functions $P_{l}^m(x)$ are the angular basis of (\ref{KGop0}). 


\par\noindent
Our aim now is to extract  the master equation in a Schr\"odinger-like form, where the latter includes the effective potential. First, we introduce the tortoise radius $r_*$ \cite{Chen:2022ynz}
\begin{equation}\label{tortoiseradius}
    \frac{dr}{dr_*}=
    {\cal F}(r)\left[1+\frac{\epsilon}{2}b^j_{lm}\left(A_j-B_j\right)\right] ,
\end{equation}
and a new radial wave function $\Psi_{l,m}$
\begin{equation}\label{Rlm}
    R_{l,m}=\frac{\Psi_{l,m}}{r}\left[1+\frac{\epsilon}{4}b^j_{lm}\left(A_j-B_j\right)-\epsilon\int dr\frac{Z_{lm}(r)}{4r^2}\right],
\end{equation}
where
\begin{equation}\label{Zim}
    Z_{lm}(r) \equiv b^j_{lm}r^2\left(A_j'-B_j'+C_j'+D_j'\right)\nonumber + 4i g^j_{lm}d_j+\frac{4i\omega r^2b^j_{lm}a_j}{\cal F}~.
\end{equation}
\par\noindent
By substituting (\ref{tortoiseradius}) and (\ref{Rlm}) into (\ref{KGequationmaster}), we obtain  the master equation written in a Schr\"odinger-like form
\begin{equation}
    \label{Schoedingerlikeform}
    \partial_{r_*}^2\Psi_{l,m}+\omega^2\Psi_{l,m}=V_{\textrm{eff}}(r)\Psi_{l,m}
\end{equation}
with  the effective potential, i.e.,  $V_{\textrm{eff}}(r)$, to be of the form \cite{Chen:2022ynz}
\begin{eqnarray}\nonumber
    V_{\textrm{eff}}(r) &=& l(l+1)\frac{{\cal F}}{r^2}+\frac{{\cal F}}{r}\frac{d{\cal F}}{dr}\left[1+\epsilon b^j_{lm}\left(A_j-B_j\right)\right]+\epsilon\Bigg\{
    \frac{{\cal F}}{r^2}\Big[a^j_{lm}\left(A_j-D_j\right)-c^j_{lm}\left(A_j-C_j\right)-\frac{b^j_{lm}}{4}\frac{d^2}{dr_*^2}\left(A_j-B_j\right)  \\
    & &- 
    \frac{d^j_{lm}}{2}\left(A_j+B_j-C_j+D_j\right)+\,e^j_{lm}\partial_r\left({\cal 
    F}c_j\right)\Big]+\frac{1}{4r^2}\frac{d}{dr_*}\left[b^j_{lm}r^2\frac{d}{dr_*}
    \left(A_j-B_j+C_j+D_j\right)\right]\Bigg\}  , \label{eq:Veff}  
\end{eqnarray}
and  with the coefficients $a^j_{lm}$, $b^j_{lm}$, $c^j_{lm}$, $d^j_{lm}$,  $e^j_{lm}$ and $g^j_{lm}$ to be given  in Appendix \ref{AppendixB}. The eikonal QNMs and photon geodesics, which form the photon sphere around black holes, are related, since in the eikonal limit, i.e., $l\gg1$ the effective potential exhibits a peak located at the photon sphere. Interestingly enough, the spacetime deformation induced by the $\Theta^2$-terms implies that the effective potential does depend on $m$ (and besides that on $l$ as in absence of deformations), affecting the behavior of high-frequency modes that could be different for different values of $m$. \yq{This splitting of frequencies for different values of the projection of angular momentum m was also reported in the QNM study of the RN black holes in noncommutative gravity \cite{Ciric:2017rnf,DimitrijevicCiric:2019hqq}.}
%
%
%
\section{Quasinormal Mode calculation in characteristic integration method}
\label{sec: QNM calculation}
%
%
%
\par\noindent
In this Section, we solve the Schr\"odinger-like equation \eqref{Schoedingerlikeform} with the effective potential \eqref{eq:Veff} utilizing the characteristic integration method. Therefore, we derive the time evolution of dominant QNM. In the following numerical calculation and plots, we choose $G=c=1$, and we choose the black hole mass as the unit of length, i.e., $M=1$. 

Considering the parameter constrain $|\Theta|<10^{-11}GeV^{-1}$ from the GUP requirement \yq{mentioned in TABLE~\ref{table1}} \cite{Kanazawa:2019llj}, for a black hole mass $M$ of around 20 mass of sun, we have $GeV^{-2}\sim10^{114}(Mc^2)^{-2}$, so that in the following discussion, we will strictly constrain our choice of parameter $\Theta$ to satisfy the requirement $\Theta^2 \ll 10^{92}(Mc^2)^{-2}$.
%
%
%
\subsection{Adding up the terms for the effective potential}
%
%
%
\par
\begin{figure}[h]
    \centering
    \includegraphics[width=0.5\textwidth]{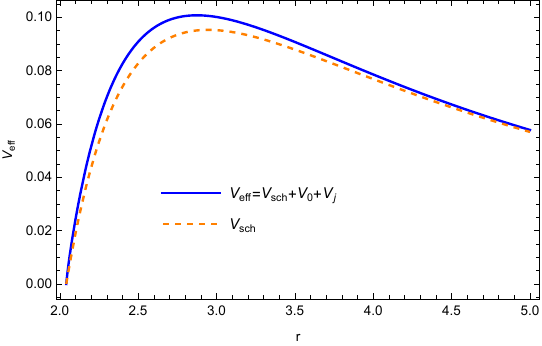}
    \caption{\it{The  effective potential $V_{\mathrm{eff}}(r)$ as a function of $r$, for $l=m=1$, and $\Theta^2=0.5$ for NC gravity.}}
    \label{fig: potential_compare with Schwarzschild}
\end{figure}

\begin{figure}[h]
    \centering
    \includegraphics[width=0.45\textwidth]{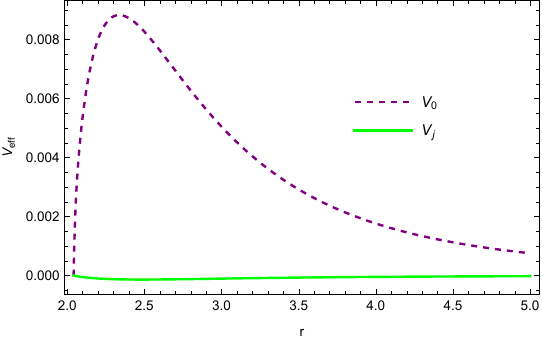}
    \includegraphics[width=0.45\textwidth]{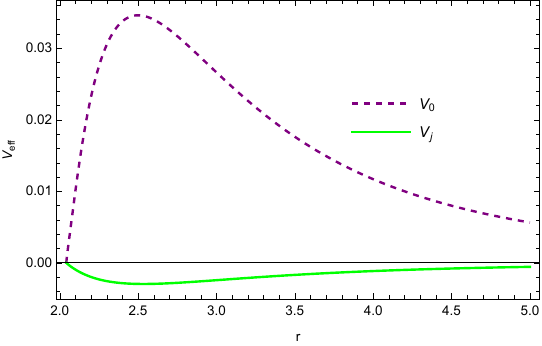}
    \caption{\it{The contribution of the $j=0$ part, i.e. $V_{0}$, as well as the contribution of the $j>0$ part, i.e. $V_{j}$, to the effective potential $V_{\mathrm{eff}}(r)$, for $\Theta^2=0.5$. The left panel is for $l=m=1$, while the right panel is for $l=3$, $m=1$. The purple dashed line is for the potential of $j=0$, and the green solid line is for the potential of $j>0$.}}
    \label{fig: potential_V0 and Vj}
\end{figure}

\noindent
We can separate the effective potential into three parts: (i) the Schwarzschild effective potential, i.e.,  $V_{\mathrm{sch}}$, (ii) the contribution from the $j=0$ part, i.e., $V_{0}$, and (iii) the contribution from the $j>0$ part, i.e., $V_{j}$,  and thus it reads
\begin{equation}
    \label{eq: effective potential}
    V_{\mathrm{eff}}(r)=V_{\mathrm{sch}}+V_{0}+V_{j},
\end{equation}
where
\begin{align}
    V_{\mathrm{sch}}=
&
    l(l+1)\frac{{\cal F}}{r^2}+\frac{{\cal F}}{r}\frac{d{\cal F}}{dr},
    \\
    V_{0}=
&
    \epsilon\frac{{\cal F}}{r}\frac{d{\cal F}}{dr} b^0_{lm}(A_0-B_0)+\epsilon\left\{\frac {{\cal F}} {r ^{2}} \left[a_{l m}^0\left(A_0-D_0\right)-c_{l m}^0\left(A_0-C_0\right)-\frac{d_{l m}^0}{2}\left(A_0+B_0-C_0+D_0\right)\right]\right.
    \notag\\
&
    \left.+\frac{1}{4 r^2} \frac{d}{d r_*}\left[b_{l m}^0 r^2 \frac{d}{d r_*}\left(A_0-B_0+C_0+D_0\right)\right]-\frac{b_{l m}^0}{4} \frac{d^2}{d r_*^2}\left(A_0-B_0\right)\right\},
    \\
    V_{j}=
&
    \label{eq: V_j}
    -\epsilon\frac{{\cal F}}{r^2}\sum_{j=1}^{\infty}\left(a^j_{lm}+\frac{1}{2}d^j_{lm}\right)D_j+\epsilon\sum_{j=1}^{\infty}\frac{1}{4r^2}\frac{d}{dr^*}\left(b^j_{lm}r^2\frac{d}{dr^*}\right)D_j~.
\end{align}
The calculation of the Schwarzschild contribution $V_{\mathrm{sch}}$ and the $j=0$ contribution $V_{0}$ is straightforward. Thus, we need to add up the terms of the $j>0$ part. As can be seen later, the integration required from the ``projection method" get divergent for the $m=0$ scenario. As a result, the ``projection method" can only be applied to the $m\neq 0$ cases to get the variable separation equations for ``quantum number " $l$. Therefore, we limit our discussion to the cases where $m\neq 0$, and give the concrete result for $l=m=1$ as an example. Calculating the coefficients $a^j_{lm}$, $b^j_{lm}$, $c^j_{lm}$, $d^j_{lm}$ with the help of MATHEMATICA, we obtain for the case of $l=m=1$
\begin{equation}
    \label{eq: a_lm}
    a_{1 1}^j=\frac{3(1+(-1)^j)}{4(1+j)},\;
    b_{1 1}^j=\frac{3[1+(-1)^j]}{2(1+j)(3+j)},\;
    c_{1 1}^j=\frac{3[1+(-1)^j](-1+j)}{4(1+j)(3+j)},\;
    d_{1 1}^j=\frac{3[1+(-1)^j]j}{2(1+j)(3+j)}~.
\end{equation}
Now, we can add up the terms of the $j>0$ part of the effective potential and acquire
\begin{equation}
    \sum_{j=1}^{\infty}\left(a^j_{1 1}+\frac{1}{2}d^j_{1 1}\right)D_j=\frac{3}{64r^2},
\end{equation}
\begin{align}
&
    \sum_{j=1}^{\infty}\frac{1}{4r^2}\frac{d}{dr^*}\left(b^j_{1 1}r^2\frac{d}{dr^*}\right)D_j
    \\
&
    =\frac{r-r_h}{65536 (r-2)^3r^{12}}\left(-56 \Theta ^2+32 r^4+3 \Theta ^2 r^3-64 r^3+24\Theta ^2 r\right)  \left(32 r^6+6 \Theta ^2r^5-64 r^5 r_h-128 r^5-6 \Theta ^2 r^4-9 \Theta ^2 r^4r_h\right.
    \notag\\
&
    \left.+256 r^4 r_h+128 r^4+96 \Theta ^2 r^3+12\Theta ^2 r^3 r_h-256 r^3 r_h-424 \Theta ^2r^2-120 \Theta ^2 r^2 r_h+448 \Theta ^2 r+528 \Theta ^2r r_h-560 \Theta ^2 r_h~\right). \notag
\end{align}
Therefore, we obtain for $l=m=1$
\begin{align}
    V_{j}=
&    \frac{r-r_h}{65536 (r-2)^4 r^{13}}( \left(131072 r^{14}+2048 r^{13} \left(3
   \Theta ^2+32 r_h-512\right)\right.
   \notag\\
&   +128 r^{12} \left(-9 \Theta
   ^4-400 \Theta ^2+16 \left(9 \Theta ^2-256\right)
   r_h+24576\right)
   \notag\\
&   +2 r^{11} \left(-27 \Theta ^6+1584
   \Theta ^4+30720 \Theta ^2+192 \left(3 \Theta ^4-272 \Theta
   ^2+4096\right) r_h-2097152\right)
   \notag\\
&   +r^{10} \left(54
   \Theta ^6-68832 \Theta ^4+458752 \Theta ^2+\left(54 \Theta
   ^6-3744 \Theta ^4+642048 \Theta ^2-2097152\right)
   r_h+2097152\right)
   \notag\\
&   +r^9 \left(\left(-81 \Theta ^6+83232
   \Theta ^4-3710976 \Theta ^2+1048576\right) r_h-3 \Theta
   ^2 \left(1287 \Theta ^4-130304 \Theta
   ^2+557056\right)\right)
   \notag\\
&   +2 \Theta ^2 r^8 \left(7155 \Theta
   ^4-665664 \Theta ^2+\left(2241 \Theta ^4-234048 \Theta
   ^2+5427200\right) r_h+1064960\right)
   \notag\\
&   -2 \Theta ^2 r^7
   \left(24 \left(1563 \Theta ^4-109280 \Theta
   ^2+20480\right)+\left(8451 \Theta ^4-796608 \Theta
   ^2+7307264\right) r_h\right)
   \notag\\
&   +4 \Theta ^2 r^6 \left(2
   \left(47529 \Theta ^2-1845952\right) \Theta ^2+\left(22005
   \Theta ^4-1545600 \Theta ^2+1847296\right) r_h\right)
   \notag\\
&   -8
   \Theta ^4 r^5 \left(131652 \Theta ^2+54963 \Theta ^2
   r_h-2151232 r_h-2545664\right)
   \notag\\
&   +32 \Theta ^4 r^4
   \left(90003 \Theta ^2+\left(37953 \Theta ^2-737600\right)
   r_h-329280\right)
   \notag\\
&   -96 \Theta ^4 r^3 \left(78160 \Theta
   ^2+\left(34269 \Theta ^2-126784\right) r_h\right)
   \notag\\
&   \left.+64
   \Theta ^6 r^2 (132753 r_h+166474)-896 \Theta ^6 r
   (13415 r_h+6342)+6397440 \Theta ^6
   r_h\right)~,
\end{align}
and, to the $\Theta^2$ order, we have
\begin{align}
    V_j(r)=
&   \frac{(r-r_h) (2 r+r_h)}{r^4}+\frac{\Theta^2}{64
   (r-2)^2 r^7} 
   \left(6 r^6+12 r^5 r_h-26 r^5-18 r^4 r_h^2-4 r^4
   r_h-68 r^4+30 r^3 r_h^2+503 r^3 r_h\right.
   \notag\\
&   \left.+280
   r^3-435 r^2 r_h^2-2044 r^2 r_h-240 r^2+1764 r
   r_h^2+2044 r r_h-1804 r_h^2\right)+\mathcal{O}\left(\Theta ^4\right)~.
\end{align}
It is evident that now we are able to compute the explicit expression of the effective potential, i.e., \eqref{eq: effective potential}, and, thus, we can plot its shape. In order to see how the effective potential of the deformed Schwarzschild black hole deviates from the effective potential of the standard Schwarzschild black hole solution, we plot both effective potentials in  FIG.~\ref{fig: potential_compare with Schwarzschild}. It is easily seen that the difference between the two effective potentials is small. This means that the contribution from the $j=0$ part, i.e., $V_{0}$, and the contribution from the $j>0$ part, i.e., $V_{j}$, are small compared to the effective potential of the standard Schwarzschild black hole solution, i.e., $V_{\mathrm{sch}}$. This is also depicted in FIG.~\ref{fig: potential_V0 and Vj} in which we plot the contribution of the $j=0$ part and the contribution of the $j>0$ part to the modified effective potential. These graphs are plotted with parameter $\Theta^2=0.5$. The left panel is ploted with $l=m=1$, while the right panel is plotted with $l=3$, $m=1$. As we can see, both $j=0$ and $j>0$ contributions are quite small, and actually the $j>0$ contribution is even smaller than the $j=0$ contribution. However, for $m=1$, as $l$ gets larger, the contribution from the $j>0$ part also gets larger, and for $l=3$ this part is not negligible compared with the contribution of the  $j=0$ part. 
\par\noindent
At this point it should be pointed out that the calculation of the explicit expression of the effective potential meets some divergence when $m=0$. In particular, for the summation that appears in the contribution of the $j>0$ part, we have an infinite summation up to $j=+\infty$. For the case of $m=0$, this summation is divergent, and, thus, we cannot get the explicit expression of the effective potential. For instance, for the case of $l=1$ and $m=0$,  combining \eqref{eq: V_j} and \eqref{eq: a_lm}, we obtain
\begin{equation}
    \sum_{j=0}^{\infty}\left(a^j_{1 0}+\frac{1}{2}d^j_{1 0}\right)D_j=\sum_{j=0}^{\infty}\frac{3(1+(-1)^j)^2 j}{64(1+j)(3+j)r^2}\sim\frac{3}{64r^2}\sum_{j=1}^{\infty}\frac{1}{j}~.
\end{equation}
\par\noindent
It is obvious that this infinite summation gives a divergent result. The same result is also obtained for the case of $l=0$ and $m=0$ as well as for the case $l=2$ and $m=0$. The conclusion is that we cannot carry out the computation for any value of $l$ when $m=0$. 
%
%
%
%
%
%
%
\par\noindent
At this point, one may think of adopting  the analysis of Ref. \cite{Chen:2022ynz} and, thus, utilize (5.17) in \cite{Chen:2022ynz} which is an approximated expression of the effective potential for the case of $m=0$. A couple of comments are in order here. First, this result does not hold in our case since we do not have the eikonal limit, i.e., $l\gg 1$. Second, if we substitute $A_{2k}=A$ and $C_{2k}=C$ in this expression, then we will find that the summation in this case is also divergent. 
\par\noindent
Finally, in our case here the divergence is generic in the sense that the explicit expression of the effective potential is divergent due to the specific background geometry, namely the deformed Schwarzschild black hole spacetime. It is known that the case $m=0$ corresponds to the polar orbits which are the orbits of the photons that pass above or nearly above the poles. However, it is easily seen in FIG.~\ref{fig: shell}, that the spacetime is not smooth at the poles due to the specific NC corrections. Therefore, polar orbits, and thus the case of $m=0$, should be excluded from the calculation of the explicit expression of the effective potential.  
%
%
%
\subsection{The numerical function for the tortoise coordinate}
%
%
%
\par
\begin{figure}[h]
    \centering
    \includegraphics[width=0.45\textwidth]{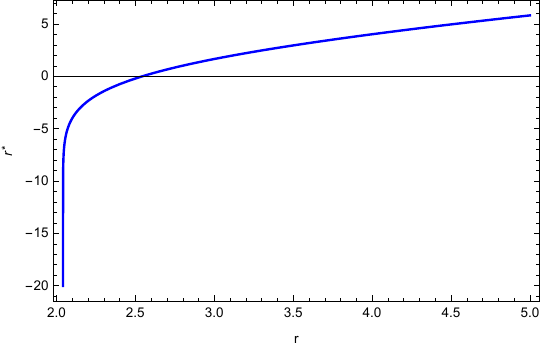}
    \includegraphics[width=0.45\textwidth]{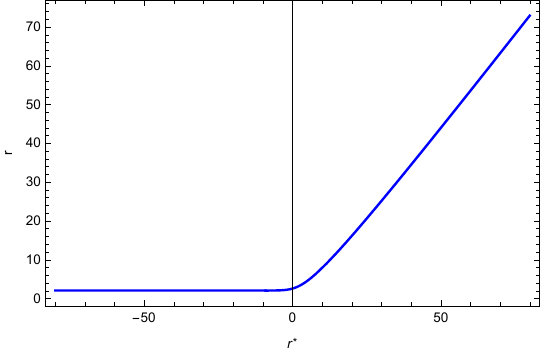}
    \caption{\it{ $r^*(r)$ as a function of $r$ (left panel), and   $r(r^*)$ 
    as function of $r^*$ (right panel), for $l=m=1$ and \yq{$\Theta^2=0.5$}.}}
    \label{fig: r2star and star2r}
\end{figure}

\noindent
The next step is to get the numerical function for the tortoise coordinate. Since we already have the function for $\frac{dr}{dr^*}$, i.e., \eqref{tortoiseradius}, we can solve the corresponding differential equation to get $r^*(r)$, and then inverse this function to get $r(r^*)$. The shape of these two functions with parameter choice $l=m=1$ and \yq{$\Theta^2=0.5$} is shown in FIG.~\ref{fig: r2star and star2r}.
%
%
%
\par\noindent

%
%
%
%
%
%
\par
\begin{figure}[h]
    \centering
    \includegraphics[width=0.5\textwidth]{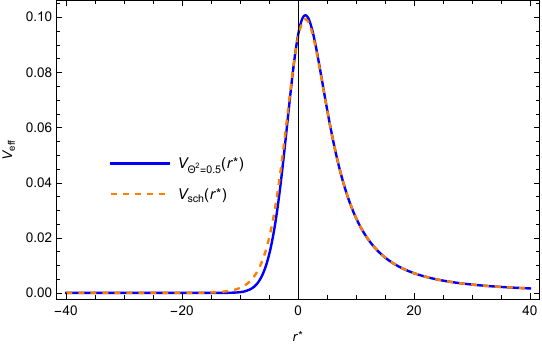}
    \caption{\it{The shape of the effective potential in terms of the tortoise coordinate $r^*$, for $l=m=1$, and $\Theta^2=0.5$.}}
    \label{fig: potential star_compare with Schwarzschild}
\end{figure}

\begin{figure}[h]
    \centering
    \includegraphics[width=0.5\textwidth]{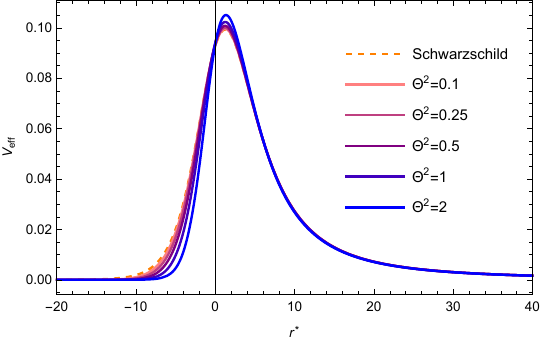}
    \caption{\it{The shape of the effective potential in terms of the tortoise coordinate $r^*$, for  $l=m=1$ and various values of $\Theta^2$.}}
    \label{fig: potential star with lots of epsilon}
\end{figure}

\noindent
In FIG.~\ref{fig: potential star_compare with Schwarzschild}, the effective potential $V_{\mathrm{eff}}$ is plotted as a function of the tortoise coordinate, namely $r^{*}$, for the case of  $l=m=1$ and $\Theta^{2}=0.5$ for NC gravity. From the plot, we can see the shape of the effective potential of the deformed Schwarzschild black hole is higher and thinner than the nondeformed Schwarzschild scenario. This can be further seen in FIG.\ref{fig: potential star with lots of epsilon} where we have plotted the effective potential $V_{\text{eff}}$ for the case of $l=m=1$ and several values of $\Theta^2$.
%
%
%
\subsection{The time evolution of the dominant mode}
%
%
%
\par\noindent
In this Section, we numerically solve the Schr$\ddot{\text{o}}$dinger-like equation utilizing the discretization method and then extract the time evolution of the dominant mode \cite{Gundlach:1993tp}. 
For convenience, first we introduce the light-cone coordinates
\begin{equation}
    u=t-r^*,\; v=t+r^{*}~.
\end{equation}
Then, the Schr$\ddot{\text{o}}$dinger-like equation \eqref{Schoedingerlikeform} can be written as
\begin{equation}
    -4\frac{\partial^2\Psi_{l,m}(u,v)}{\partial u\partial v}-V_\text{eff}(u(r),v(r))\Psi_{l,m}(u,v)=0~.
\end{equation}
Discretizing this equation, we obtain
\begin{equation}
    \Psi_{l,m}(N)=\Psi_{l,m}(W)+\Psi_{l,m}(E)-\Psi_{l,m}(S)-\frac{h^2}{8}V_\text{eff}(S)(\Psi_{l,m}(W)+\Psi_{l,m}(E))+\mathcal{O}(h^4),
\end{equation}
with $S=(u,v)$, $W=(u+h,v)$, $E=(u,v+h)$, $N=(u+h,v+h)$. In this discretization method, the step length is set to $h=0.1$ while the initial and boundary condition is set on the null boundary $u=u_0$ and $v=v_0$.
\begin{figure}[h]
    \centering
    \includegraphics[width=0.35\textwidth]{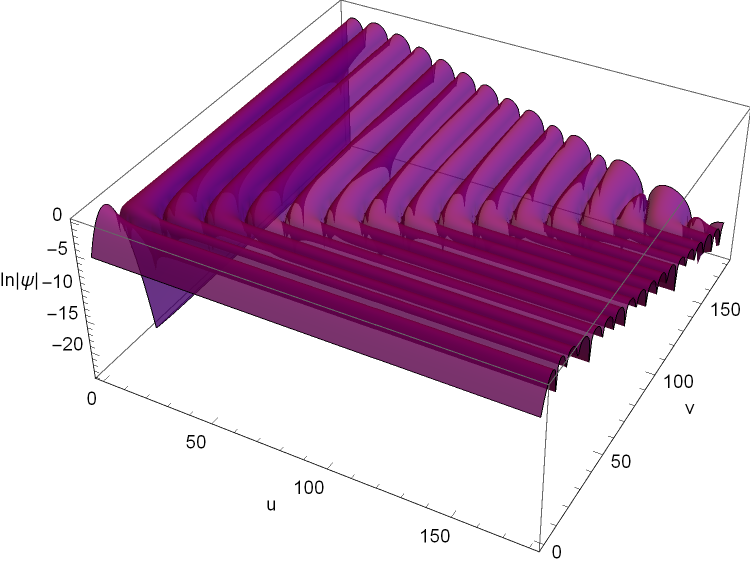}
    \includegraphics[width=0.55\textwidth]{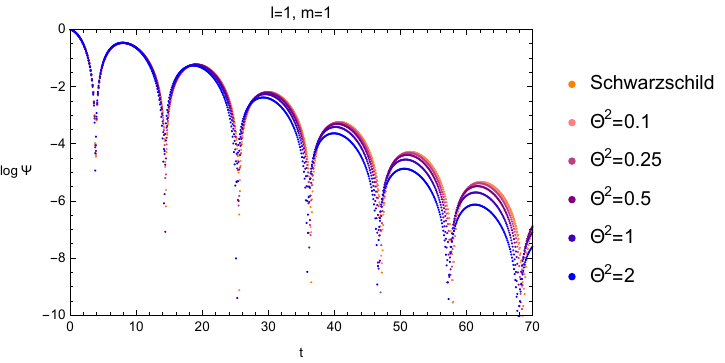}
    \caption{\it{The evolution of the mode function of the massless test scalar field for the case   $l=m=1$. The left panel shows the 3D evolution of the mode function with $\Theta^2=0.5$. The right panel shows the time evolution of the mode function on constant radius $r^*$.}}
    \label{fig: QNM evolution}
\end{figure}
\par\noindent
In FIG.~\ref{fig: QNM evolution}, we present an example of the evolution of the mode function of the massless test scalar field for the case of $l=m=1$. In the left panel, we show the 3D evolution of the massless test scalar field for the case of $\Theta^2=0.1$. In the right panel, we show the time evolution of mode function located at constant $r^*$. 
Lines with different colors in the figure stand for different values of the non-commutative parameter $\Theta$, as shown in the legend.

The estimation of QNM frequencies through nonlinear function fitting of the time evolution is given in TABLE. \ref{tab: WKB_epsilon}. WKB method is a widely used semi-analytical method for QNM calculations~\cite{Iyer:1986np,Molina:2003dc}. For comparison, in TABLE.\ref{tab: WKB_epsilon} we also give the result from WKB method with a $P^6_6$ Pade approximation~\cite{Matyjasek:2017psv}, which is typically 2 orders of magnitude better than the ordinary WKB approximation. From the table we can see the result from the two methods have the same overall tendency, but don't exactly agree with each other because the precision of the nonlinear fitting of the time evolution data is quite limited, especially for the small $l$ case.

\begin{table}
    \begin{tabular}{ccc}
        \hline
         $\Theta^2$ & $\omega(\text{WKB})$ & $\omega(\text{evolution})$\\
         \hline
         0 \; & 0.306-0.101 i\; & 0.280-0.104 i  \\
         0.1\; & 0.294-0.0983 i\; & 0.291-0.091 i  \\
         0.25\; & 0.294-0.0990 i\; & 0.292-0.092 i  \\
         0.5\; & 0.295-0.1012 i\; & 0.293-0.095 i \\
         1.0\; & 0.297-0.1059 i\; & 0.295-0.099 i  \\
         2.0\; & 0.295-0.1155 i\; & 0.295-0.107 i  \\
         \hline
    \end{tabular}
    \caption{WKB calculation and time evolution estimation results for the QNM frequencies of different noncommutative parameter $\Theta^2$ and $l=m=1$. The WKB result is derived through WKB method with a $P^6_6$ Pade approximation. The time evolution result is derived from nonlinear function fitting.}
    \label{tab: WKB_epsilon}
\end{table}
%
%
%
\section{Conclusions}
\label{Conclusions}
%
%
%
\par\noindent
In this work we have calculated the QNM frequencies of a
 massless test scalar field around  static black hole solutions in noncommutative gravity.  
As a first step we obtained the master equation, which is of a Schr$\ddot{\text{o}}$dinger-like form and, thus, the effective potential was explicitly written. The effective potential is made of three parts: (i) the Schwarzschild effective potential, i.e.  $V_{\mathrm{sch}}$, (ii) the contribution from the $j=0$ part, i.e. $V_{0}$, and (iii) the contribution from the $j>0$ part, i.e. $V_{j}$. In order for these parts to be computed,  we excluded the polar orbits ($m=0$) in order to avoid a divergence due to the fact that the deformed Schwarzschild black hole is ``broken" at the poles. Furthermore,  a tortoise coordinate was employed in order to make the computation of the effective potential easier, and, additionally, a discretization method was utilized.

Using several diagrams, we have shown that noncommutative gravity does have an effect on  the quasinormal modes evolution compared to the one obtained in the framework of GR.

Our results are relevant in the perspective of increasing accuracy in observations of gravitational waves from binary gravitational systems, whose characteristic oscillation modes can provide  interesting information. 
Furthermore, the analysis performed in this paper could be used as an additional tool to test the GR predictions and examine whether gravitational modifications of the specific form induced by noncommutative geometry are possible.\\
\yq{We end with a couple of comments. 
First, in our analysis we have computed the scalar QNM frequencies and not the `real' QNM frequencies of a static black hole, since we solved the Klein-Gordon equation of the scalar particle as opposed to the Einstein's equations of GR.
Therefore, one may question if strictly speaking, our 
work is relevant to real GWs. 
However, the point is that our equations 
take a Schr\"odinger-like form and, which has many 
similarities, and practically, is of the same form as the Einstein's equations for GWs in GR, up to polarization factors. Therefore, we expect the bulk of our conclusions to hold for real GWs. 
Second, we note that there have been works computing QNM frequencies in the framework of NC spacetimes \cite{Herceg:2023zlk,Herceg:2023pmc}, in the framework of $f(R)$ gravity \cite{Suvorov:2019qow}, as well as in the framework of LQG \cite{Cruz:2015bcj,Villani:2020bfc, Bouhmadi-Lopez:2020oia,Cruz:2020emz,Yang:2023gas}.}
%
%
%
\section*{Acknowledgements}
This work was supported by the Natural Sciences and
Engineering Research Council of Canada. ENS acknowledges participation in the COST 
Association Action CA18108 ``{\it Quantum Gravity Phenomenology in the 
Multimessenger Approach (QG-MM)}''. 
YFC acknowledges the support by National Key R\&D Program of China (2021YFC2203100), by NSFC (11961131007, 11653002, 12261131497), by Fundamental Research Funds for Central Universities, by CSC Innovation Talent Funds, by USTC Fellowship for International Cooperation, by USTC Research Funds of the Double First-Class Initiative, by CAS project for young scientists in basic research (YSBR-006). 
All numerics were operated on the computer 
clusters {\it LINDA} \& {\it JUDY} in the particle cosmology group at USTC.
%
%
%
\appendix
%
%
%
\section{The first-order operator $\mathcal{D}_{(1)m,\omega}^2$}
\label{AppendixA}
%
%
%
\par\noindent
The first order operator  $\mathcal{D}_{(1)m,\omega}^2$ in (\ref{KGop}) is given by \cite{Chen:2022ynz}
\begin{widetext}
\begin{align}
\mathcal{D}_{(1)m,\omega}^2&=\frac{m^2{\cal F}}{r^2\sin^2\theta}\left(A_j-D_j\right)\cos^j\theta-\frac{{\cal F}}{r^2}\left(A_i-B_j\right)\cos^j\theta\left[\partial_r\left(r^2{\cal F}\partial_r\right)\right]-\frac{{\cal F}^2}{2}\left(A_j'-B_j'+C_j'+D_j'\right)\cos^j\theta\partial_{r}\nonumber\\&-\frac{{\cal F}}{r^2}\left(A_j-C_j\right)\cos^j\theta\left(\cot\theta\partial_\theta+\partial_\theta^2\right)-\frac{{\cal F}}{2r^2}\left[\left(A_j+B_j-C_j+D_j\right)\partial_\theta\cos^j\theta\right]\partial_\theta-\frac{2i\omega {\cal F}}{r}a_j\cos^j\theta\left(r\partial_r+1\right)\nonumber\\&-i\omega {\cal F}\partial_ra_j\cos^j\theta-\frac{2i\omega}{r^2}b_j\cos^j\theta\partial_\theta-\frac{i\omega}{r^2\sin\theta}b_j\partial_\theta\left(\cos^j\theta\sin\theta\right)-\frac{im{\cal F}}{r^2\sin^2\theta}\left[2d_j\cos^j\theta {\cal F}\partial_r+\partial_r\left({\cal F}d_j\right)\cos^j\theta\right]\nonumber\\
&+\frac{im{\cal F}}{r^4\sin^3\theta}e_j\left(j\cos^{j-1}\theta\sin^2\theta+\cos^{j+1}\theta\right)-\frac{2im{\cal F}}{r^4\sin^2\theta}e_j\cos^j\theta\partial_\theta\nonumber\\
&+\frac{{\cal F}}{r^2}\left[\partial_{r}\left({\cal F}c_j\right)\cos^j\theta\partial_\theta+2{\cal F}c_j\cos^j\theta\partial_{r\theta}^2\right]+\frac{{\cal F}^2}{r^2\sin\theta}c_j\partial_\theta\left(\cos^j\theta\sin\theta\right)\partial_r\,.\label{KGop1}
\end{align}
\end{widetext}
The summations over $j$ are implicitly assumed in each term.
%
%
%
\section{The coefficients of the effective potential (\ref{eq:Veff}) }
\label{AppendixB}
%
%
%
\par\noindent
The coefficients $a^j_{lm}, ...., h^j_{lm}$ of the effective potential 
(\ref{eq:Veff})  are given as \cite{Chen:2022ynz}
\begin{align}
a^j_{lm}&=\frac{m^2}{\mathcal{N}_{lm}}\int_{-1}^1\frac{x^j\left({P_l^m}\right)^2}{1-x^2}dx\,,\\
b^j_{lm}&=\frac{1}{\mathcal{N}_{lm}}\int_{-1}^1x^j\left({P_l^m}\right)^2dx\,,\\
c^j_{lm}&=\frac{1}{\mathcal{N}_{lm}}\int_{-1}^1x^j{P_l^m}\left[\left(1-x^2\right)\partial_x^2-2x\partial_x\right]{P_l^m}dx\,,\\
d^j_{lm}&=\frac{1}{\mathcal{N}_{lm}}\int_{-1}^1{P_l^m}\left(1-x^2\right)\left(\partial_xx^j\right)\left(\partial_x{P_l^m}\right)dx\,,\\
e^j_{lm}&=\frac{-1}{\mathcal{N}_{lm}}\int_{-1}^1dxx^j{P_l^m}\sqrt{1-x^2}\partial_x{P_l^m}\,,\\
f^j_{lm}&=\frac{1}{\mathcal{N}_{lm}}\int_{-1}^1dx\left({P_l^m}\right)^2\left[\frac{x^{j+1}}{\sqrt{1-x^2}}-\sqrt{1-x^2}\partial_xx^j\right]\,,\\
g^j_{lm}&=\frac{m}{\mathcal{N}_{lm}}\int_{-1}^1\frac{x^j\left({P_l^m}\right)^2dx}{1-x^2}\,,\\
h^j_{lm}&=\frac{m}{\mathcal{N}_{lm}}\int_{-1}^1\frac{\left({P_l^m}\right)^2dx}{\left(1-x^2\right)^{3/2}}\left[jx^{j-1}\left(1-x^2\right)+x^{j+1}\right]+\frac{2m}{\mathcal{N}_{lm}}\int_{-1}^1\frac{x^j{P_l^m}\left(\partial_x{P_l^m}\right)dx}{\sqrt{1-x^2}}
\end{align}
with
\begin{equation}
\int_{-1}^1dxP^m_l(x)P^m_k(x)=\frac{2(l+m)!}{(2l+1)(l-m)!} \, \delta_{lk}\,,
\end{equation}
and
\begin{equation}
    \mathcal{N}_{l m} \equiv \frac{2(l+m) !}{(2 l+1)(l-m) !}.
\end{equation}
%
%
%
\bibliography{reference}
\bibliographystyle{apsrev4-1}
%
%
%

\end{document}